\documentstyle[aps,preprint,prl,citesort,epsf]{revtex}

\newcommand{\eqref}[1]{Eq.~(\protect\ref{#1})}

 \tightenlines

\begin{document}

\draft

\title{Extensive Scaling and Nonuniformity of the Karhunen-Lo\`eve 
Decomposition for the Spiral-Defect Chaos State}

\author{Scott M. Zoldi$^{\dagger}$, 
Jun Liu$^{\triangle}$, Kapil M. S. Bajaj$^{\ast}$,
Henry S. Greenside$^{\dagger\dagger}$, and Guenter Ahlers$^{\ast}$}


\address{$^{\dagger}$ Theoretical Division and Center for 
Nonlinear Studies, Los Alamos National Laboratory, 
Los Alamos, New Mexico 87545}
\address{$^{\triangle}$ Schlumberger - ATE, M/S 930, 1601 
Technology Dr., San Jose, California 95110}
\address{$^{\ast}$Department of Physics and Center for Nonlinear Science,
University of California, Santa Barbara, California 93106}
\address{$^{\dagger\dagger}$Department of Physics and Center for Nonlinear and
Complex Systems, Duke University, Durham, North Carolina 27708}

\date{\today}
\maketitle


\begin{abstract}
By analyzing large-aspect-ratio spiral-defect-chaos (SDC)
convection images, we show that the 
Karhunen-Lo\`eve decomposition (KLD) scales extensively for 
subsystem-sizes larger than $4d$ ({\it d} is the fluid depth), which
strongly suggests that SDC is extensively chaotic.  From this
extensive scaling, the intensive length~$\xi_{\rm KLD}$ is
computed and found to have a different dependence on
the Rayleigh number than the two-point correlation length~$\xi_{\rm
2}$.  Local computations of~$\xi_{\rm KLD}$ reveal a substantial
spatial nonuniformity of SDC that extends over radii $18d< r < 45d$ in 
a $\Gamma=109$ aspect-ratio cell.
\end{abstract}


           

\narrowtext

A significant theoretical challenge is to find ways to characterize
the nonperiodic time-dependent patterns often observed in large
sustained nonequilibrium systems~\cite{Cross93}.  In a recent
paper~\cite{ZG97}, a new length scale for characterizing
spatiotemporal chaos (STC), the KLD length~$\xi_{\rm KLD}$, was
proposed based on the {\em extensive scaling} of the Karhunen-Lo\`eve
decomposition (KLD)~\cite{Lumley96}.  This length scale was
shown~\cite{ZG97} to be computable from moderate amounts of space-time
data and to contain information similar to the fractal-dimension
density, which has not yet been computed from experimental
data~\cite{density-refs}.  For certain idealized mathematical models,
the length~$\xi_{\rm KLD}$ was shown to have a different parametric
dependence than the commonly computed two-point correlation
length~$\xi_{2}$ and so provides a different way to characterize
spatiotemporal chaos~\cite{ZG97}. Further, this length~$\xi_{\rm KLD}$
could be calculated from data localized to a region of space and so
offered a way to analyze spatial inhomogeneities.  However, an
application of the KLD length to {\em experimental} data had not been
made prior to this work.

In this paper, we present the first application of the KLD length to
experimental STC data by analyzing the recently discovered~\cite{MBCA}
spiral-defect chaos (SDC) in Rayleigh-B\'{e}nard convection.  We
analyze the SDC state of large aspect-ratio ($\Gamma \equiv r/d =
29~{\rm and}~109$ where $r$ and $d$ are the radius and thickness of
the cell) cells and find that the KLD dimension~$D_{\rm KLD}$ of the
data scales extensively with subsystem volume when the diameter of the
subvolume is larger than about 4{\it d} where {\it d} is the depth of
the fluid.  This extensivity strongly suggests that the SDC state is
extensively chaotic \cite{density-refs,extensive-comment}, which
provides the first such evidence for an experimental system.  From the
extensive scaling of the dimension~$D_{\rm KLD}$ with subvolume, we
calculate the length~$\xi_{\rm KLD}$ which reflects the density of
linearly independent modes needed to approximate the spatiotemporal
data~\cite{ZG97}.  For the reduced Rayleigh numbers~$\epsilon$
explored in this study,~$\xi_{\rm KLD}$ exhibits a minimum with
increasing~$\epsilon$ while the two-point correlation length~$\xi_{\rm
2}$ monotonically decreases.  We speculate that this different
behavior in the KLD length arises from changes in the structure of 
spirals and straight rolls with increasing~$\epsilon$.  Finally, we also
show that the length~$\xi_{\rm KLD}$ provides a way to measure a
dynamical inhomogeneity in SDC data within the interior of the cell.
A local KLD analysis demonstrates that SDC is {\em not} dynamically
uniform (Most numerical simulations of SDC~\cite{theory} have used
periodic lateral boundary conditions, for which the dynamics should be
statistically homogeneous by translational invariance).  In addition,
this nonuniformity extends to within a radius $r=18d$ of the center of
an aspect-ratio $\Gamma = 109$ cell.  These results suggest that the
length~$\xi_{\rm KLD}$ can provide useful insights about
spatiotemporal chaos.

We tested the utility of the length~$\xi_{\rm KLD}$ for analyzing
SDC experimental data~\cite{MBCA,SDCmore} 
by collecting thousands of shadowgraph images~\cite{experiment} of the SDC
state from a Rayleigh-B\'{e}nard experiment using compressed CO$_2$
gases with Prandtl number $\sigma \approx 1$ and in cylindrical cells
of aspect ratio $\Gamma = 29$ and 109.  SDC appeared for 
$\epsilon \equiv {\Delta T \slash \Delta T_c} - 1$ above 0.56 and 
0.23 respectively for the $\Gamma = 29$ and 109 cells, where $\Delta T_c$ 
is the critical temperature difference for the onset of convection.
Five sets of SDC data were taken in
the two cells as summarized in Table~\ref{datasets}. We chose
sampling times $\Delta t$ that were significantly longer than
the correlation time $\tau \sim 10 t_v$ (where $t_v$ is the vertical 
thermal diffusion time) to maximize the amount 
of uncorrelated data~\cite{MBCA}.  Good spatial resolution of the
convection patterns required at least 5 pixels per convection roll for 
the $\Gamma=109$ cell.  Data was taken from the central 43\% of the 
$\Gamma=109$ convection cell, as indicated 
in~Fig.~\protect\ref{fig:convection-pictures}.

To illustrate the KLD analysis used to study the experimental data,
let~$u(t_i,{\bf x}_j)$ denote the light intensity of an
experimental image at position~${\bf x}_j$ at time~$t_i$.  The analysis
proceeds by constructing a $T \times S$ space-time matrix of data,
\begin{equation}
{A_{ij}}~=~u(t_i,{\bf x}_j)-<u(t_i,{\bf x}_j)>
\end{equation}
where $<u(t_i,{\bf x}_j)>$ denotes the time-average (average over index $i$) 
of the set of
experimental images~$u(t_i,{\bf x}_j)$, $T$ is the number of
observation times~$t_i$, and $S$ is the number of observation
sites~${\bf x}_j$.  The KLD dimension~$D_{\rm KLD}$~\cite{Sirovich89} of 
the matrix $A_{ij}$ then measures the number of linear eigenmodes needed to
approximate some fraction~$0<f<1$~of the variance of the experimental
data and can be computed from the eigenvalues of the matrix~${\bf
A}^{T} {\bf A}$~\cite{ZG97,Sirovich89,kldrefs}.  We compute $D_{\rm
KLD}({\bf x_j})$ for concentric subsystems of volume~$V$ (square or
circular geometry) that are centered at a particular point $\bf x_j$ in space.
$D_{\rm KLD}({\bf x_j})$ depends on the point~$\bf x_j$ and so
provides a measure of dynamical inhomogeneity. If $D_{\rm KLD}({\bf
x_j})$ increases linearly with subsystem volume~$V$ and with a slope
$\delta$, then the length~$\xi_{\rm KLD}$ is defined to be
$\delta^{-1/{\rm d}}$ where d is the dimensionality of the data (${\rm d}=2$
for SDC).

In applying these ideas to SDC data, we computed $D_{\rm KLD}$ for a
fixed fraction~$f=0.7$~\cite{fraction-comment} and for larger and
larger square subimages in the center of the convection cell of size
$S \times S$, where~$2d<S<25d$.  As shown in
Fig.~\protect\ref{fig:extensive-scaling}, the dimension~$D_{\rm KLD}$
scales approximately linearly with subsystem data over a relatively
large range of subsystem sizes~$4d<S<13d$ provided that a
sufficiently long time series was used.  This extensive scaling,
together with the arguments in Ref~\cite{ZG97} relating $\xi_{\rm
KLD}$ to the dimension correlation length $\xi_{\delta}$, strongly
suggests that the SDC state is extensively chaotic.  We note that the 
extensive linear scaling of the dimension~$D_{\rm
KLD}$ with subsystem area is best for smaller subsystems, which we
believe is a consequence of the fact that smaller subsystems have a
faster time scale to become statistically stationary.

We next compared $\xi_{\rm KLD}$ (computed in the center of the cell 
for $f=0.7$) with the two-point correlation length~$\xi_2$ as 
the reduced Rayleigh number~$\epsilon$ was varied.  The two-point
correlation length~$\xi_{\rm 2}$ was calculated from the inverse of
the width of the peak in the Fourier spectrum of the spatial data 
which was pre-multiplied with a hanning window of diameter equal to 
the lateral dimension of the images.  
As shown in Table \ref{lengthscales}, the
parametric dependences of the length~$\xi_{\rm KLD}$ and the two-point
correlation length~$\xi_{\rm 2}$ are different~\cite{fraction-comment}.  
The fact that~$\xi_{\rm
KLD}$ attains a minimum between~$\epsilon=0.79~{\rm and}~0.88$, 
and then increases with
increasing~$\epsilon$ (corresponding to a decrease in complexity) is
somewhat counterintuitive since one might have expected $\xi_{\rm KLD}$ to
decrease with~$\epsilon$ as the system is forced further away from
equilibrium (more modes are needed to approximate the space-time
data).  A possible explanation for these opposing trends may be
found by examining the different spatial structures in
Figs.~\protect\ref{fig:convection-pictures}A 
and~\protect\ref{fig:convection-pictures}B.  For
Fig.~\protect\ref{fig:convection-pictures}A ($\epsilon=0.52$), the
area fraction of local straight-roll regions is larger than that for
spiral-roll regions, whereas for
Fig.~\protect\ref{fig:convection-pictures}B ($\epsilon=0.93$) the
relationship between straight- and spiral-roll regions is reversed.
We speculate that the data between~$\epsilon=0.79~{\rm and}~0.88$ 
consist of nearly equal
fractions of straight- and spiral-roll regions and thus require more
KLD eigenmodes per unit volume~$D_{\rm KLD}/V_{\rm sub}$ and so~$\xi_{\rm
KLD}=(D_{\rm KLD}/V_{\rm sub})^{-1/2}$ is smaller.  This speculation
about the relative complexity of straight-roll versus spiral-roll regions
is supported by examining the KLD spatial eigenmodes for the~$\epsilon
=0.52$ and $\epsilon=0.93$ data.  These modes have the same
qualitative symmetries up to eigenmode 21~($f$=0.45 for $\epsilon=0.52$), 
beyond which the local straight-roll regions in the 
$\epsilon=0.52$ data entered into the next KLD spatial modes and 
broadened the KLD eigenvalue spectrum\cite{ZG97}.  In the $\epsilon=0.93$ 
data, beyond eigenmode 21 the eigenmodes consisted of lattices of 
convection rolls.

As the KLD analysis of a subsystem is a local procedure, one can
quantify differences in the SDC in different regions of the convection
cell and so test the assumption that SDC is dynamically homogeneous.
For the $\Gamma=109$ cell, we investigated a radial dependence of
$\xi_{\rm KLD}$ by first establishing that $D_{\rm KLD}$ scaled
extensively for an annulus of radius~$r_o<r<r_o+d$ and
angular sector that varied from~$\Delta \theta=0$ to~$\Delta
\theta=2\pi$ radians (extensivity was with respect to~$\Delta \theta$).  
We could not estimate nonuniformity for~$r_o <
8d$ because the subsystems were too small, or for $r_o > 45d$ as this
exceeded the area imaged by the CCD camera.
Fig.~\protect\ref{fig:kld-radial} shows how the length $\xi_{\rm KLD}$
increases by 10$\%$ from $r=8d$ to $r=45d$.  The variation in the
length $\xi_{\rm KLD}$ with radial distance $r$ demonstrates that SDC
cannot be considered homogeneous~(Fig.~\protect\ref{fig:kld-radial})
even in large aspect ratio cells.  Based on this calculation we
consider the cell {\em approximately} uniform for $r<18d$.
The time-average of
the SDC data does not resemble the nonuniformity in the KLD length in
the interior of the cell, but the more easily computed variance
pattern does resemble that of $\xi_{\rm KLD}$.  
The 10\% nonuniformity of $\xi_{KLD}$ in the large cell may be due to both
dynamical and experimental reasons.  The dynamical one would be due to the
pinning of the pattern by the side-wall 
which has been shown to cause time-averaged
patterns near the sidewall for instance in rotating convection 
experiments~\cite{Ning93}. (We also found nonunformity of $\xi_{KLD}$ 
near the sidewall in the $\Gamma = 29$ cell.)  The experimental source 
could come from the nonlinearity of shadowgraph method.   For thin cells and 
high epsilon values, the nonlinearity is strong and can make the 
shadowgraph method sensitive to small optical nonuniformities.  
Unfortunately, the relative strength of these two sources of 
nonuniformity cannot be determined.  

In summary, we have carried out the first analysis of experimental
data using the extensive scaling of KLD in subsystems~\cite{ZG97} for
the spiral-defect chaos state.  KLD analysis is straightforward to
apply to small subsystems of different geometry as it does not impose
an approximate periodicity of the space-time data as is the case for
Fourier analysis.  
By verifying that the dimension~$D_{\rm KLD}$
scaled linearly with subsystem size, we provide strong evidence that
the experimental system is extensively chaotic~\cite{Cross93}.  The
utilization of a subsystem was essential for our study for two
reasons.  First, local analysis of subsystems allowed the
characterization of inhomogeneous dynamics in our experimental
convection cell.  Further, by exploiting the reduced data requirements
of subsystems, we could estimate the parametric behavior of the
dimension density (the average number of degrees of freedom per unit
area) in the center of the convection cell.  For the SDC data
analyzed, the length~$\xi_{\rm KLD}$ seems to quantify differences in
the fraction of straight-roll and spiral-defect regions (also observed
in the KLD spatial eigenmodes), and so provides information beyond that
available from~$\xi_{\rm 2}$.  Finally the nonuniformity
of the KLD length was shown to extend over 80$\%$ of the radius in
the~$\Gamma=109$ convection cell.  
Future analysis will hopefully 
provide further insights about the transition to SDC and about the 
chiral-symmetry breaking recently observed in SDC for rotating 
convection experiments~\cite{future}.

S.M.Z was supported in part by the Computational
Graduate Fellowship Program of the Office of
Scientific Computing in the Department of Energy.
S.M.Z and H.S.G were supported 
by NSF grants NSF-DMS-93-07893 and NSF-CDA-92123483-04, 
and by DOE grant DOE-DE-FG05-94ER25214. J.L, K.B, and G.A were
supported by DOE grant DOE-DE-FG03-87ER13738.


\bibliographystyle{prsty}  



\begin{references}
\bibitem{extensive-comment}
The strict definition of extensive chaos is that the fractal dimension
scales extensively with system volume~\cite{Cross93}.  However the 
KLD dimension~$D_{\rm KLD}$ has been shown to scale extensively with
system volume when the fractal dimension scales extensively~\cite{ZG97}.
\bibitem{fraction-comment} 
The specific value of $D_{\rm KLD}$ depends on the fraction $f$ but the
parametric dependence of $D_{\rm KLD}$ is often weakly dependent on
the choice of $f$~\cite{ZG97}.  The fraction~$f$ can not be chosen 
too close to 1.0 as this would include
errors in the experimental data arising from 
variations in the intensity of the shadowgraph images and from 
discretization error.
\end{references}

\newpage
\begin{figure}   
\caption{Time-dependent shadowgraph patterns for the central~43\% of 
the~$\Gamma=109$ 
convection  cell, with light and dark regions corresponding to cool 
and warm fluid respectively. (A) and (B) are snap-shots of the SDC pattern 
for reduced Rayleigh number~$\epsilon=0.52$ and~$\epsilon=0.93$
respectively.}
\label{fig:convection-pictures}
\end{figure}

\begin{figure}   
\caption{Scaling of the KLD dimension~$D_{\rm KLD}$  with 
subsystem area~$N^{2}$ for data in the central~43\% of the~$\Gamma = 109$ 
cell with reduced Rayleigh number~$\epsilon=0.52$ and variance
fraction~$f=0.70$.  The area~$N^{2}$  
of a subsystem is measured in units of d where d is the 
the thickness of the cell.  
The labels indicate the number of images used in the calculations and
lines connecting points were drawn to guide the eye.}
\label{fig:extensive-scaling}
\end{figure}

\begin{figure}   
\caption{
Radial inhomogeneities of the length~$\xi_{\rm KLD}$ in annular 
subsystems of fixed radii and increasing azimuthal angle for 
the~$\Gamma = 109$ convection cell ($\epsilon=0.79$).  The
radial distance r is measured in units of d (the thickness of the
convection cell) from the center of the $\Gamma=109$ convection
cell.
}
\label{fig:kld-radial}
\end{figure}

\newpage
\epsfysize=8.5in{\epsfbox{figsdc1.epsi}}

\newpage
\centerline{\epsfysize=8.5in \epsfbox{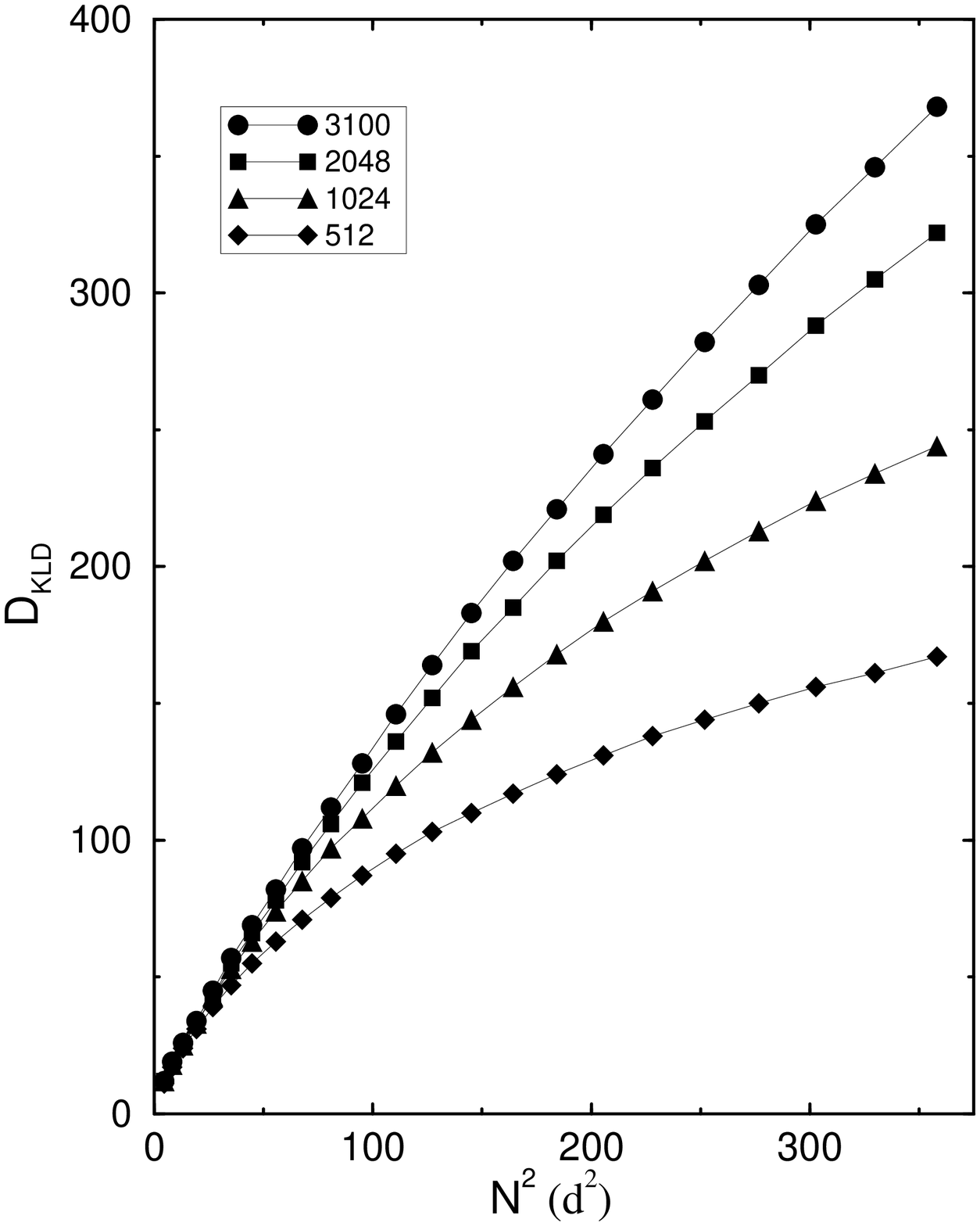}}

\newpage
\epsfysize=4.5in \epsfbox{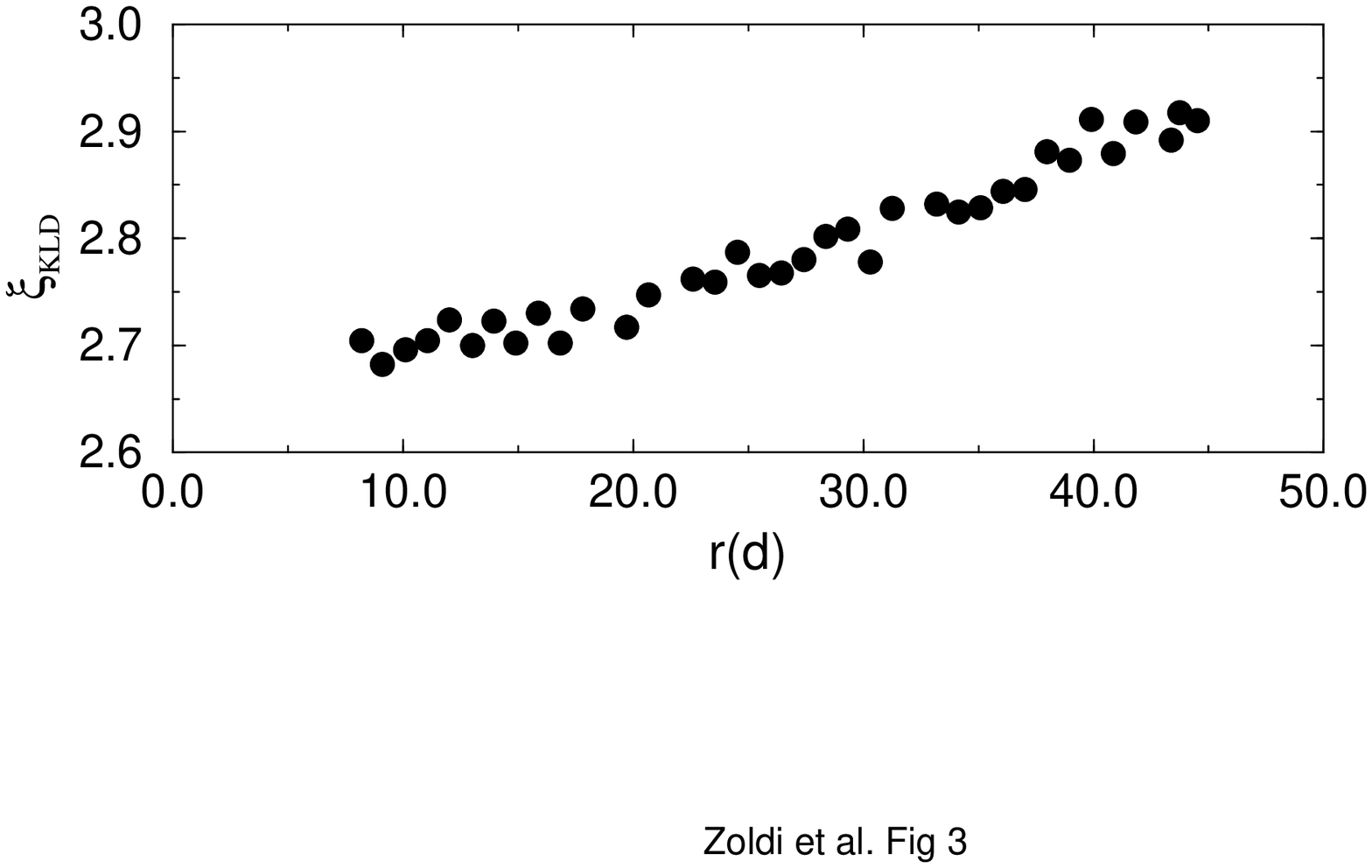}
\twocolumn
\begin{table}
\caption{Parameters for data used to calculate $\xi_{\rm KLD}$.
  $\Gamma$ is the aspect ratio, $\epsilon$ is the reduced Rayleigh
number, $T$ is the number of images, and $\Delta t/t_v$ is the
sampling rate in units of the vertical thermal diffusion time $t_v$.}
\vskip 0.1in
\begin{tabular}{cdccdd}
   $\Gamma$ & $\epsilon$ & $T$ & $\Delta t/t_v$ & $t_v$ (s) \\
\tableline
29 & 1.80 & 2048 & 33 & 8.28 \\
109 & 0.52 & 3100 & 1050 & 0.86 \\
109 & 0.67 & 2500 & 1050 & 0.86 \\
109 & 0.79 & 2500 & 1050 & 0.86 \\
109 & 0.93 & 2570 & 728 & 0.82 \\
\end{tabular}
\label{datasets}
\end{table}

\begin{table}
\caption{Lengths $\xi_{\rm KLD}$ and $\xi_{2}$ 
(normalized to the depth of the fluid $d$)
as functions of
aspect ratio $\Gamma$ and reduced Rayleigh number $\epsilon$.  
The length $\xi_{\rm KLD}$ was calculated for fraction $f=0.7$.}
\vskip 0.1in
\begin{tabular}{ddcd}
   $\Gamma$ & $\epsilon$ & $\xi_{\rm KLD}$/d & $\xi_{\rm 2}$/d \\
\tableline
29  & 1.80  & 1.47 & 2.28  \\
109 & 0.52 & 0.84 & 4.38 \\
109 & 0.67 & 0.81 & 3.85 \\
109 & 0.79 & 0.75 & 3.60 \\
109 & 0.88 & 0.74 & 3.39 \\
109 & 0.93 & 0.90 & 3.22 \\
\end{tabular}
\label{lengthscales}
\end{table}

\end{document}